\documentclass[aps,prl,twocolumn,superscriptaddress,showpacs]{revtex4}

\usepackage{graphicx} 
\usepackage{dcolumn}  
\usepackage{bm}       
\usepackage{amssymb}

\newcommand{\GeVCSq}{\ensuremath{\GeV\!/c^2}}
\newcommand{\GeV}{\ensuremath{\mathrm{Ge\kern -0.1em V}}}
\newcommand{\ppbar} {\ensuremath{p\overline{p}}}
\newcommand{\TeV}{\ensuremath{\mathrm{Te\kern -0.1em V}}}
\newcommand {\sigmaBr}{\ensuremath{\sigma \cdot {\cal B}}}
\newcommand{\Zg}{{\it Z}$/\gamma^*$ } 
\newcommand {\sigmaBrXee}{\ensuremath{\sigma \cdot {\cal B}(X \rightarrow ee)}}
\newcommand {\sigmaBrGee}{\ensuremath{\sigma \cdot {\cal B}(G \rightarrow ee)}}

\begin{document}

\title{Search for New Physics in High Mass Electron-Positron Events in $\ppbar$ Collisions at $\sqrt{s} = 1.96$~\TeV}

\affiliation{Institute of Physics, Academia Sinica, Taipei, Taiwan 11529, Republic of China} 
\affiliation{Argonne National Laboratory, Argonne, Illinois 60439} 
\affiliation{Institut de Fisica d'Altes Energies, Universitat Autonoma de Barcelona, E-08193, Bellaterra (Barcelona), Spain} 
\affiliation{Baylor University, Waco, Texas  76798} 
\affiliation{Istituto Nazionale di Fisica Nucleare, University of Bologna, I-40127 Bologna, Italy} 
\affiliation{Brandeis University, Waltham, Massachusetts 02254} 
\affiliation{University of California, Davis, Davis, California  95616} 
\affiliation{University of California, Los Angeles, Los Angeles, California  90024} 
\affiliation{University of California, San Diego, La Jolla, California  92093} 
\affiliation{University of California, Santa Barbara, Santa Barbara, California 93106} 
\affiliation{Instituto de Fisica de Cantabria, CSIC-University of Cantabria, 39005 Santander, Spain} 
\affiliation{Carnegie Mellon University, Pittsburgh, PA  15213} 
\affiliation{Enrico Fermi Institute, University of Chicago, Chicago, Illinois 60637} 
\affiliation{Comenius University, 842 48 Bratislava, Slovakia; Institute of Experimental Physics, 040 01 Kosice, Slovakia} 
\affiliation{Joint Institute for Nuclear Research, RU-141980 Dubna, Russia} 
\affiliation{Duke University, Durham, North Carolina  27708} 
\affiliation{Fermi National Accelerator Laboratory, Batavia, Illinois 60510} 
\affiliation{University of Florida, Gainesville, Florida  32611} 
\affiliation{Laboratori Nazionali di Frascati, Istituto Nazionale di Fisica Nucleare, I-00044 Frascati, Italy} 
\affiliation{University of Geneva, CH-1211 Geneva 4, Switzerland} 
\affiliation{Glasgow University, Glasgow G12 8QQ, United Kingdom} 
\affiliation{Harvard University, Cambridge, Massachusetts 02138} 
\affiliation{Division of High Energy Physics, Department of Physics, University of Helsinki and Helsinki Institute of Physics, FIN-00014, Helsinki, Finland} 
\affiliation{University of Illinois, Urbana, Illinois 61801} 
\affiliation{The Johns Hopkins University, Baltimore, Maryland 21218} 
\affiliation{Institut f\"{u}r Experimentelle Kernphysik, Universit\"{a}t Karlsruhe, 76128 Karlsruhe, Germany} 
\affiliation{High Energy Accelerator Research Organization (KEK), Tsukuba, Ibaraki 305, Japan} 
\affiliation{Center for High Energy Physics: Kyungpook National University, Taegu 702-701, Korea; Seoul National University, Seoul 151-742, Korea; SungKyunKwan University, Suwon 440-746, Korea} 
\affiliation{Ernest Orlando Lawrence Berkeley National Laboratory, Berkeley, California 94720} 
\affiliation{University of Liverpool, Liverpool L69 7ZE, United Kingdom} 
\affiliation{University College London, London WC1E 6BT, United Kingdom} 
\affiliation{Centro de Investigaciones Energeticas Medioambientales y Tecnologicas, E-28040 Madrid, Spain} 
\affiliation{Massachusetts Institute of Technology, Cambridge, Massachusetts  02139} 
\affiliation{Institute of Particle Physics: McGill University, Montr\'{e}al, Canada H3A~2T8; and University of Toronto, Toronto, Canada M5S~1A7} 
\affiliation{University of Michigan, Ann Arbor, Michigan 48109} 
\affiliation{Michigan State University, East Lansing, Michigan  48824} 
\affiliation{University of New Mexico, Albuquerque, New Mexico 87131} 
\affiliation{Northwestern University, Evanston, Illinois  60208} 
\affiliation{The Ohio State University, Columbus, Ohio  43210} 
\affiliation{Okayama University, Okayama 700-8530, Japan} 
\affiliation{Osaka City University, Osaka 588, Japan} 
\affiliation{University of Oxford, Oxford OX1 3RH, United Kingdom} 
\affiliation{University of Padova, Istituto Nazionale di Fisica Nucleare, Sezione di Padova-Trento, I-35131 Padova, Italy} 
\affiliation{LPNHE, Universite Pierre et Marie Curie/IN2P3-CNRS, UMR7585, Paris, F-75252 France} 
\affiliation{University of Pennsylvania, Philadelphia, Pennsylvania 19104} 
\affiliation{Istituto Nazionale di Fisica Nucleare Pisa, Universities of Pisa, Siena and Scuola Normale Superiore, I-56127 Pisa, Italy} 
\affiliation{University of Pittsburgh, Pittsburgh, Pennsylvania 15260} 
\affiliation{Purdue University, West Lafayette, Indiana 47907} 
\affiliation{University of Rochester, Rochester, New York 14627} 
\affiliation{The Rockefeller University, New York, New York 10021} 
\affiliation{Istituto Nazionale di Fisica Nucleare, Sezione di Roma 1, University of Rome ``La Sapienza," I-00185 Roma, Italy} 
\affiliation{Rutgers University, Piscataway, New Jersey 08855} 
\affiliation{Texas A\&M University, College Station, Texas 77843} 
\affiliation{Istituto Nazionale di Fisica Nucleare, University of Trieste/\ Udine, Italy} 
\affiliation{University of Tsukuba, Tsukuba, Ibaraki 305, Japan} 
\affiliation{Tufts University, Medford, Massachusetts 02155} 
\affiliation{Waseda University, Tokyo 169, Japan} 
\affiliation{Wayne State University, Detroit, Michigan  48201} 
\affiliation{University of Wisconsin, Madison, Wisconsin 53706} 
\affiliation{Yale University, New Haven, Connecticut 06520} 
\author{T.~Aaltonen}
\affiliation{Division of High Energy Physics, Department of Physics, University of Helsinki and Helsinki Institute of Physics, FIN-00014, Helsinki, Finland}
\author{A.~Abulencia}
\affiliation{University of Illinois, Urbana, Illinois 61801}
\author{J.~Adelman}
\affiliation{Enrico Fermi Institute, University of Chicago, Chicago, Illinois 60637}
\author{T.~Affolder}
\affiliation{University of California, Santa Barbara, Santa Barbara, California 93106}
\author{T.~Akimoto}
\affiliation{University of Tsukuba, Tsukuba, Ibaraki 305, Japan}
\author{M.G.~Albrow}
\affiliation{Fermi National Accelerator Laboratory, Batavia, Illinois 60510}
\author{S.~Amerio}
\affiliation{University of Padova, Istituto Nazionale di Fisica Nucleare, Sezione di Padova-Trento, I-35131 Padova, Italy}
\author{D.~Amidei}
\affiliation{University of Michigan, Ann Arbor, Michigan 48109}
\author{A.~Anastassov}
\affiliation{Rutgers University, Piscataway, New Jersey 08855}
\author{K.~Anikeev}
\affiliation{Fermi National Accelerator Laboratory, Batavia, Illinois 60510}
% ...
\collaboration{CDF Collaboration\footnote{With visitors from $^a$University of Athens, 15784 Athens, Greece, 
$^b$University of Bristol, Bristol BS8 1TL, United Kingdom, 
$^c$University Libre de Bruxelles, B-1050 Brussels, Belgium, 
$^d$Cornell University, Ithaca, NY  14853, 
$^e$University of Cyprus, Nicosia CY-1678, Cyprus, 
$^f$University College Dublin, Dublin 4, Ireland, 
$^g$University of Edinburgh, Edinburgh EH9 3JZ, United Kingdom, 
$^h$University of Heidelberg, D-69120 Heidelberg, Germany, 
$^i$Universidad Iberoamericana, Mexico D.F., Mexico, 
$^j$University of Manchester, Manchester M13 9PL, England, 
$^k$Nagasaki Institute of Applied Science, Nagasaki, Japan, 
$^l$University de Oviedo, E-33007 Oviedo, Spain, 
$^m$University of London, Queen Mary College, London, E1 4NS, England, 
$^n$University of California Santa Cruz, Santa Cruz, CA  95064, 
$^o$Texas Tech University, Lubbock, TX  79409, 
$^p$University of California, Irvine, Irvine, CA  92697, 
$^q$IFIC(CSIC-Universitat de Valencia), 46071 Valencia, Spain, 
}}
\noaffiliation


\date{\today}

\begin{abstract}

We report the results of a search for a narrow resonance in electron-positron events in the invariant mass range of $150-950$~\GeVCSq\ using 1.3~fb$^{-1}$ of \ppbar\ collision data at $\sqrt{s} = 1.96$~\TeV\ collected by the CDF~II detector at Fermilab. No significant evidence of such a resonance is observed and we interpret the results to exclude the standard model-like $Z'$ with a mass below 923 \GeVCSq\ and the Randall-Sundrum graviton with a mass below 807 \GeVCSq\ for $k/\bar{M}_{pl}$=0.1, both at the 95\% confidence level. Combining with di-photon data excludes the Randall-Sundrum graviton for masses below 889 \GeVCSq\ for $k/\bar{M}_{pl}$=0.1.

\end{abstract}


\maketitle

At hadron colliders, electron-positron pairs ($ee$) are a distinct experimental signature with a low background rate. Since many models introducing new physics beyond the standard model of particle physics (SM) predict an excess in $ee$ production at a hadron collider, this channel has a strong discovery potential.
This Letter describes a search for a new high mass state in $ee$ events from $p\bar{p}$ collisions
at $\sqrt{s}=1.96$~TeV. The data used in this analysis were collected by the CDF~II detector at the Fermilab Tevatron and correspond to an integrated luminosity of 1.3~fb$^{-1}$. The analysis also uses results from the $\gamma\gamma$ channel described in~\cite{diPhoPaper} to increase the analysis's sensitivity to the Randall-Sundrum (RS) graviton~\cite{rsgrav}. 

The search is optimized for new physics processes which produce narrow $ee$ resonances~\cite{narRes}, but is otherwise model-independent. In addition to the above search, cross section times branching ratio (\sigmaBr) upper limits~\cite{95Limits} are set for generic neutral spin-1 and spin-2 bosons. These \sigmaBr\ limits are then used to set lower bounds on the masses of specific particles predicted by new physics models. These particles are the $E_6$ $Z'$s~\cite{zPrimeRef} and the RS graviton.
The $E_6$ model unifies the forces of the SM into the $E_6$ gauge group and in doing so predicts the presence of two additional neutral massive spin-1 bosons, referred to as $Z'$s, which can mix with some arbitrary mixing angle. The $Z'_\eta$, $Z'_\chi$, $Z'_\psi$, and $Z'_I$ $Z'$s correspond to specific values of the mixing angle and are used to benchmark the model.  
The RS graviton is predicted by the RS model of warped extra dimensions which solves the hierarchy between the weak and Planck scales by introducing an extra spatial dimension with negative curvature $k$. 
The model predicts a series of narrow neutral spin-2 resonances which couple to all SM particles, with the lowest mass resonance referred to here as the RS graviton. The properties of this model are determined by the mass of the RS graviton and the ratio $k/\bar{M}_{pl}$, where $\bar{M}_{pl}$ is the reduced effective Planck scale. This ratio governs the couplings of the graviton to SM particles and it has a favored range of 0.01 to 0.1~\cite{kMplRange}.

 The most recent similar search by the D\O\ collaboration used an integrated luminosity of 260~pb$^{-1}$ and treated the $ee$ and the di-photon ($\gamma\gamma$) channels as a single channel to perform the first dedicated RS graviton search \cite{d0RSGrav}. The most recent search for new physics in the $ee$ channel by the CDF collaboration was a dedicated $Z'$ search and used an integrated luminosity of 448~pb$^{-1}$~\cite{448paper}.

This analysis is based on an integrated luminosity of 1.3~fb$^{-1}$ collected with the CDF~II detector. The CDF~II detector is a general purpose detector which is azimuthally and forward-backward symmetric and is described in detail elsewhere~\cite{CDF}. The relevant components for this analysis are the central tracking chamber (COT) and the central and plug calorimeters. The COT is a 96--layer drift chamber placed within a 1.4~Tesla magnetic field and is used to measure the momenta of charged particles within the pseudorapidity range $|\eta| \leq 1.1$~\cite{cdfCoordinate}. The COT is complemented by a silicon microstrip detector which directly surrounds the beampipe and has a tracking coverage of $|\eta| \leq 2.0$~\cite{silicon}. The central and plug calorimeters are sampling calorimeters that surround the COT; they consist of electromagnetic (EM) and hadronic sections that measure the energy of particles in the range $|\eta| \leq 1.1$ and $1.2\leq |\eta| \leq 3.6$ respectively.

The trigger used in this analysis requires two separate deposits of EM energy in the calorimeter and is effectively 100\% efficient for selecting $ee$ events within the acceptance of the analysis.
Events are selected by requiring two electron candidates with $E_T \geq 25$~\GeV. 
Events are separated into two channels: central-central (CC) where both electrons are in the central EM calorimeter (CEM) ($|\eta| \leq 1.1$) and central-plug (CP) where one electron is in the plug EM calorimeter (PEM) ($1.2\leq |\eta| \leq 3.0$).
Electrons in the CEM are required to have a well--measured track, whereas there is no tracking requirement for electron candidates in the PEM. Due to the lack of tracking in the plug region there is no opposite sign requirement for any of the $ee$ pairs. 
In the CC channel, 5\% of the electron pairs in the signal region ($M_{ee} \geq 150$~\GeVCSq) are same sign, which is compatible with the fraction of misidentified opposite sign pairs predicted by simulation. Electrons are identified in an identical way to the previously published analysis~\cite{448paper}, with the exception that a photon conversion veto is applied to central electrons in CP events. This selection cut improves the sensitivity of the analysis by reducing the $\gamma\gamma$ background in this channel. 
The event selection and the search method defined later were chosen without regard to events observed in the signal region to ensure a statistical robust result.

The geometric and kinematic acceptance as a function of resonance mass is estimated using event samples generated by Monte Carlo (MC) simulation. The {\sc pythia} event generator~\cite{pythia}, with the {\sc CTEQ5L} parton distribution functions (PDF)~\cite{CTEQ5L} and the CDF II detector simulation based on {\sc geant}~3~\cite{GEANT} are used to generate all simulation samples unless otherwise stated. A $Z'$ with the couplings of the SM $Z$ (SM-like $Z'$) is used for the simulated spin-1 signal sample and a RS graviton with $k/\bar{M}_{pl} = 0.1$ is used for the spin-2 sample. Both the $Z'$ and RS graviton bosons are constrained to be within $\pm10\%$ of their on-shell mass at generator level. 
The uncertainty on the acceptance resulting from the PDF parameterization is estimated to be 2-4\% using the procedure recommended by the CTEQ Collaboration \cite{cteqPdfErrs}. The uncertainty on the acceptance due to initial state radiation (ISR) is estimated to be 4\% by varying the parameters governing ISR in {\sc pythia}. 
The electron identification efficiency ranges from $90\%$ to $95\%$ per electron and is estimated using the simulated signal samples. These estimates are corrected for imperfections in the simulation by comparing the simulation with the data at the $Z$ pole, resulting in a 2\% systematic uncertainty. The total selection efficiency $\times$ acceptance over the entire signal region is in the range of 40\% to 45\% for spin-1 and spin-2 bosons.  

\begin{figure}[htbp]
\begin{center}
\includegraphics[width=8.8cm,height=6.35cm,clip,trim = 0mm 00mm 0mm 12.5mm]{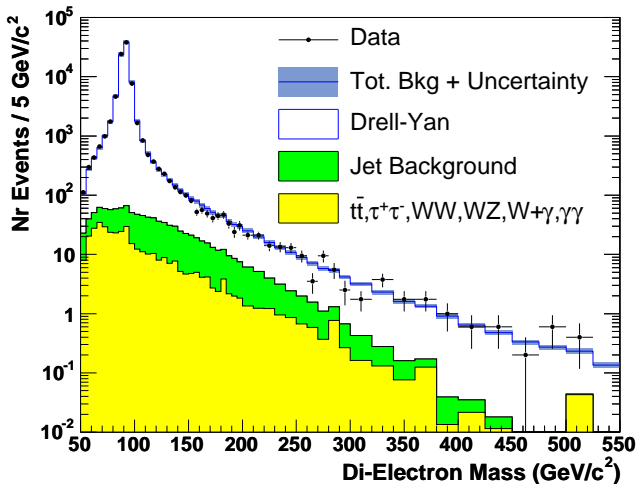}
\caption{The measured $ee$ mass spectrum with the expected background for the CC and CP channels combined. The backgrounds are displayed cumulatively. There are no observed events above 550 \GeVCSq.}
\label{fig:mass}
\end{center}
\end{figure}

The most significant source of background to new physics in the $ee$ channel is the SM Drell-Yan process via \Zg which is an irreducible background. Jet events, such as di-jet or $W+$jet events where the jets are misidentified as electrons, represent the most significant reducible background. 
Other less significant backgrounds result from $t\bar{t}$, $\tau^+\tau^-$, $WW$, $WZ$, $W\gamma$, and $\gamma\gamma$ events. Figure~\ref{fig:mass} shows the invariant mass distribution for all the background components properly normalized, together with the observed data for the CC and CP channels combined. 

The SM Drell-Yan contribution is estimated using MC simulated events normalized to the data at the $Z$ pole. By investigating the stability of the normalization factor a 4\% systematic uncertainty is obtained on the SM Drell-Yan normalization.   
An uncertainty on the SM Drell-Yan shape due to PDF uncertainties is determined using the same method as used for the acceptance. 
The di-jet and $W+$jet backgrounds are treated as a single background, referred to as the jet background. 
The size of this background is estimated from a sample of jet events constructed from the data identically to the signal sample except that at least one electron candidate is not isolated~\cite{isol} and therefore likely to be a jet. From the distribution of the isolation of these `electron' candidates, the number of jet events in the signal sample and its uncertainty is extracted.
The shape of the jet background is estimated from a jet$+Y$ sample, where $Y$ is either an electron or a jet misidentified as an electron. 
In the region where the jet background is significant, the normalization uncertainty is the dominant uncertainty on the jet background.
Using these methods, the jet background is estimated to account for $0.8\pm0.7\%$ and $25\pm8\%$ of the total background above 150 \GeVCSq\ in the CC and CP channels respectively. 
The remaining backgrounds are all estimated using MC simulation normalized to the theoretical next-to-leading order (NLO) or higher order cross section. 
The uncertainties in these background estimates are dominated by the 6\% uncertainty on the luminosity~\cite{lumiErr}. 

\begin{figure}[htbp]
\begin{center}
\includegraphics[width=8.8cm,height=6.35cm,clip,trim = 0mm 00mm 0mm 13mm]{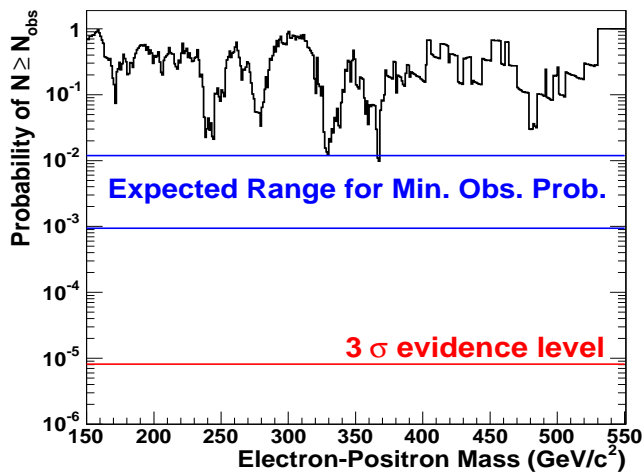}
\caption{The probability that the background alone could give rise to the observed number of events in a mass window equal to the width of a narrow resonance in the CDF detector. The expected range and the $3\sigma $ evidence line are defined in the text. 
The region $550-950$ \GeVCSq\ is not shown as no events are observed in data and therefore the p-value is always 1 in this region. }
\label{fig:prob}
\end{center}
\end{figure}

A model--independent search for an excess over SM predictions is performed in an invariant mass range of $150-950$~\GeVCSq. The search is optimized for a narrow resonance, but still retains sensitivity to other signals which would produce an excess over SM predictions.
Using 1~\GeVCSq\ steps from $M_{ee} = 150$ to 950~\GeVCSq, the probability, referred to as the p-value, of observing at least as many events as recorded in the real experiment given the expected background rate is calculated using Poisson statistics in a mass window of $4.8 + 0.044\times M_{ee}$~\GeVCSq. This mass window is approximately the width a narrow resonance would have if observed in the CDF detector, and this choice of mass window maximizes the sensitivity to discovering such a resonance for this particular analysis, as verified from studies of simulated events. 
The uncertainty on the background estimate is treated as a nuisance parameter with a Gaussian distribution.
The p-values for the CC and CP channels are combined multiplicatively and the results are shown in Fig.~\ref{fig:prob}. 
The minimum p-value expected in the absence of new physics depends on the size of the search range, with the expected minimum p-value decreasing as the search range increases. The range in which minimum p-value is expected to occur is shown in Fig. 2 and is defined to include $68.3\%$ of the minimum p-values centered on the median value, using one million simulated mass spectra.
Similarly, the 3$\sigma$ evidence line is the p-value above which the minimum p-value in 99.85\% of the simulated mass spectra fall; any p-value lower than this would be taken as evidence for the presence of new physics. The lowest p-value observed is at 367~\GeVCSq\ and is within the expected range. It is therefore concluded that the results of this analysis are consistent with the SM.

\begin{figure}[htbp]
\begin{center}
\includegraphics[width=8.8cm,height=6.35cm,clip,trim = 0mm 00mm 0mm 13mm]{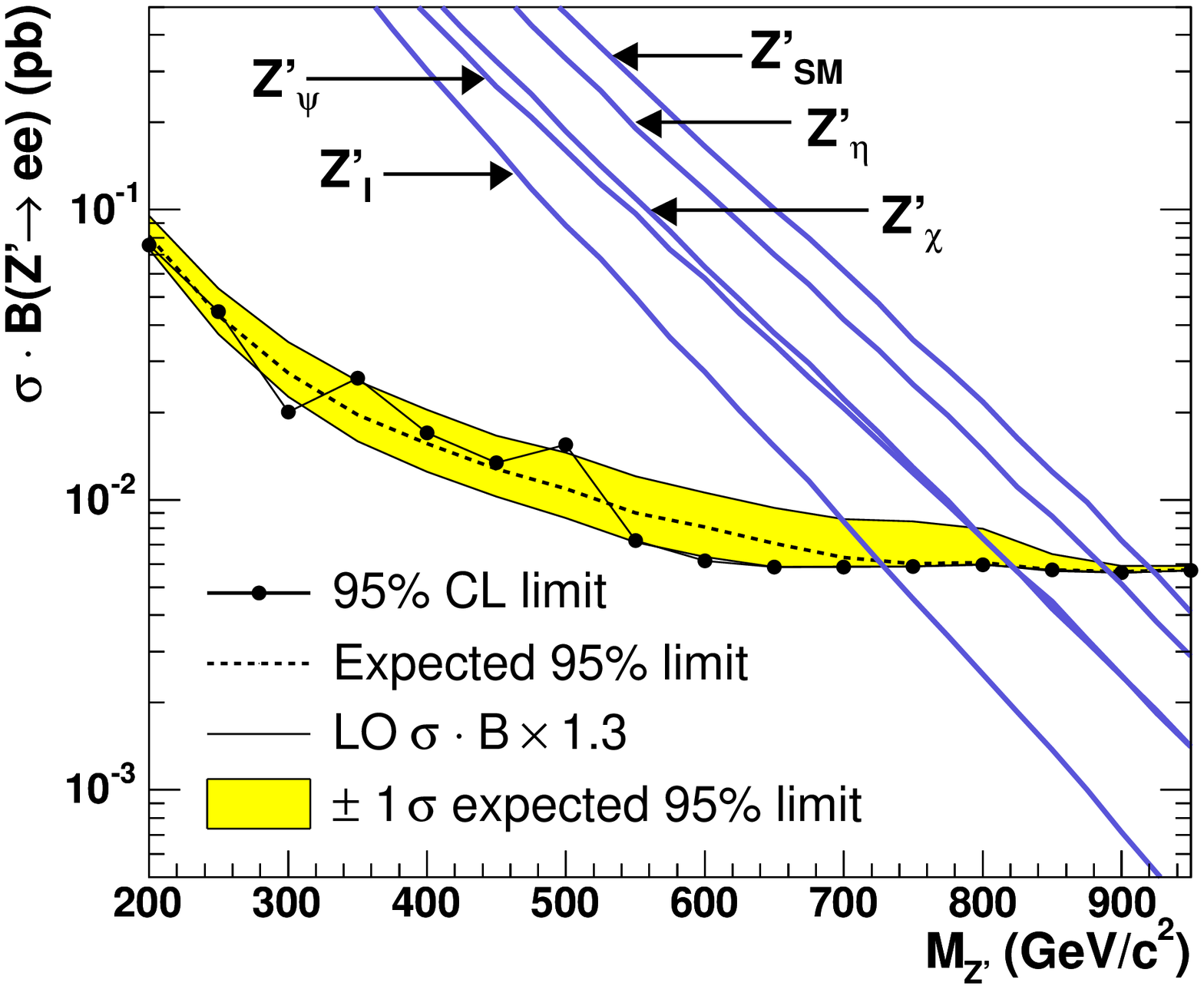}
\caption{The observed and expected limits on the \sigmaBrXee\ of a spin-1 particle.}
\label{fig:limits}
\end{center}
\end{figure}

To complement the above search, a Bayesian binned likelihood method is used to extract limits on \sigmaBrXee, where the mass of $X$ is within $\pm10\%$ of its on-shell mass. As the acceptance of the final-state $ee$ system is required to extract a cross section, it is necessary to specify the spin of the particle. Both spin-1 and spin-2 particles are considered here. The likelihood is 1-dimensional with the signal cross section as the free parameter and the bin contents treated using Poisson statistics. The likelihood is then convoluted with a Gaussian to allow for the uncertainty on the cross section from the acceptance, background, and luminosity estimates. 
The probability density function is formed by taking a flat prior for the signal cross section and is numerically integrated to obtain the limit on \sigmaBrXee.
The observed limits are shown in Fig.~\ref{fig:limits} for the spin-1 case. 
The $Z'$ model lines are obtained using the leading-order {\sc pythia} event generator using the couplings in~\cite{e6couplings}, with a factor of 1.3 applied multiplicatively to account for NLO corrections~\cite{kfactor}.

For the specific case of the RS graviton, which has a branching ratio to $\gamma\gamma$ twice that of $ee$, the analysis sensitivity can be improved by combining with the $\gamma\gamma$ channel described in~\cite{diPhoPaper}. The $\gamma\gamma$ events are required to have two photons with $E_T \geq 15$~\GeV, with one in the CEM ($|\eta|\leq1.04$) and the other either in the CEM or PEM ($1.2\leq|\eta|\leq 2.8$). The CC and CP channels use integrated luminosities of 1.2~fb$^{-1}$ and 1.1~fb$^{-1}$ respectively, with the difference arising from the requirement that the silicon system be operational for track veto purposes for photons in the PEM. Selected photons have similar isolation and shower shape requirements to electrons; however photons are required not to have an associated track, which ensures there is no overlap between the $\gamma\gamma$ and $ee$ sample.

\begin{figure}[htbp]
\begin{center}
\includegraphics[width=8.8cm,height=12.1cm,trim = 0mm 10mm 0mm 22mm, clip]{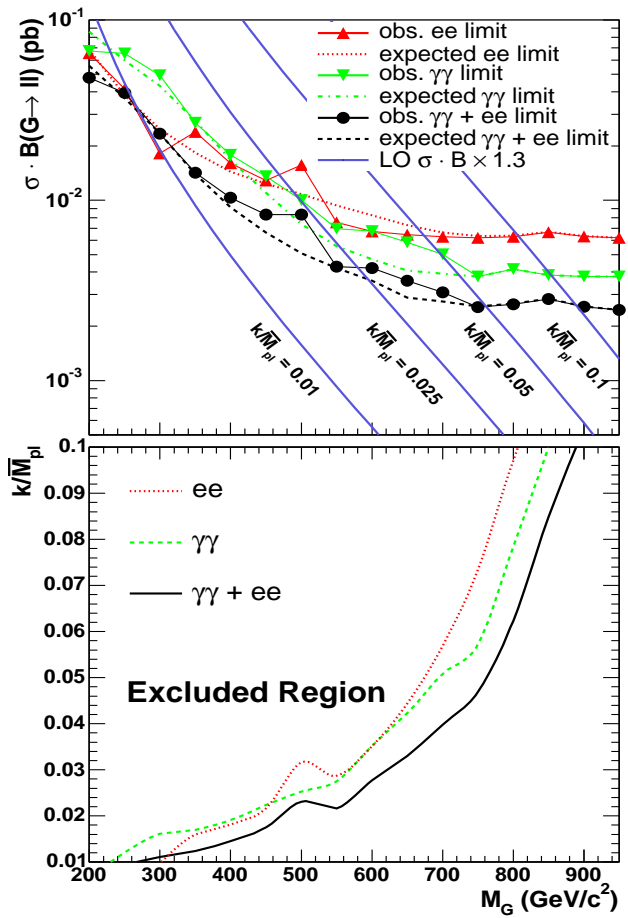}
\caption{The limits on \sigmaBrGee\ (top) and $k/\bar{M}_{pl}$ (bottom) for a RS graviton in the $ee$, $\gamma\gamma$ channels separately and combined.}
\label{fig:diPhoDiEleLimits}
\end{center}
\end{figure}

The $\gamma\gamma$ channel is combined with the $ee$ channel by multiplying their individual likelihoods together. The uncertainties on the background estimates are considered to be uncorrelated. The uncertainties on the acceptance and luminosity are taken to be 100\% correlated. The combined limits, together with the $k/\bar{M}_{pl}$ vs graviton mass exclusion region, are shown in Fig.~\ref{fig:diPhoDiEleLimits}. The RS graviton model lines are obtained using the {\sc herwig} event generator \cite{herwig} with a factor of 1.3 applied multiplicatively to account for NLO corrections. 

In summary a search has been made for new physics in the $ee$ channel, and no significant excess over the standard model prediction is observed. Limits are placed on new spin-1 and spin-2 bosons. The SM-like $Z'$ is found to be excluded for masses below 923~\GeVCSq\ and the $E_6$ $Z'$ bosons; the $Z'_I$, the $Z'_\psi$, the $Z'_\chi$, and the $Z'_\eta$ bosons are excluded with masses below 729, 822, 822, and 891~\GeVCSq\ respectively. The direct limits presented here on all the $E_6$ $Z'$ bosons surpass the corresponding indirect limits from LEP~\cite{lepLimits}. The RS graviton with $k/\bar{M}_{pl} = 0.1$ is excluded for masses below 807~\GeVCSq. When combined with the $\gamma\gamma$ channel, masses less than 889~\GeVCSq\  are excluded for $k/\bar{M}_{pl} = 0.1$.
The above limits on $Z'$s and the RS graviton represent the best single--experiment direct limits to date.

\begin{acknowledgments}
We thank the Fermilab staff and the technical staffs of the participating institutions for their vital contributions. This work was supported by the U.S. Department of Energy and National Science Foundation; the Italian Istituto Nazionale di Fisica Nucleare; the Ministry of Education, Culture, Sports, Science and Technology of Japan; the Natural Sciences and Engineering Research Council of Canada; the National Science Council of the Republic of China; the Swiss National Science Foundation; the A.P. Sloan Foundation; the Bundesministerium f\"ur Bildung und Forschung, Germany; the Korean Science and Engineering Foundation and the Korean Research Foundation; the Science and Technology Facilities Council and the Royal Society, UK; the Institut National de Physique Nucleaire et Physique des Particules/CNRS; the Russian Foundation for Basic Research; the Comisi\'on Interministerial de Ciencia y Tecnolog\'{\i}a, Spain; the European Community's Human Potential Programme; the Slovak R\&D Agency; and the Academy of Finland.
\end{acknowledgments}


\begin{thebibliography}{99}
\bibitem{narRes}
    A narrow resonance is defined as a resonance where the resolution of the CDF detector solely determines the observed width.
\bibitem{95Limits}
    All limits presented in this Letter are at the 95\% confidence level.
\bibitem{cdfCoordinate}
    CDF uses a cylindrical coordinate system with the $z$-axis along the proton direction, where $z$=0 is the center of the detector. Pseudorapidity defined as $\eta = - \ln(\tan(\theta/2))$, where $\theta$ is the polar angle while $E_T = E \sin(\theta)$, where $E$ is the energy deposited in the cluster of calorimeter towers.
\bibitem{isol}
    Electrons are isolated if in a surrounding cone of 0.4 radius the $E_T$ deposited minus 1.02 $\times$ the electron $E_T$ is less than 3~\GeV (1.6~\GeV) in the CEM (PEM).
\end{thebibliography}
\end{document}